# Elucidating the Behavior of Nanophotonic Structures Through Explainable Machine Learning Algorithms

Christopher Yeung[1,3], Ju-Ming Tsai[1], Brian King[1], Yusaku Kawagoe[1,2], David Ho[1], Mark Knight[3], and Aaswath P. Raman[1,*]

[1]*Department of Materials Science and Engineering, University of California, Los Angeles, CA 90024, USA*
[2]*Department of Mechanical Engineering, Nagoya University, Nagoya, Aichi 464-8601, Japan*
[3]*Northrop Grumman Corporation, Redondo Beach, CA 90278 , USA*
*\*Corresponding author:* aaswath@ucla.edu

**ABSTRACT:** A central challenge in the development of nanophotonic structures is identifying the optimal design for a target functionality, and understanding the physical mechanisms that enable the optimized device's capabilities. Previously investigated design methods for nanophotonic structures, including both conventional optimization approaches as well as nascent machine learning (ML) strategies, have made progress, yet they remain 'black boxes' that lack explanations for their predictions. Here we demonstrate that convolutional neural networks (CNN) trained to predict the electromagnetic response of classes of metal-dielectric-metal metamaterials, including complex freeform designs, can be explained to reveal deeper insights into the underlying physics of nanophotonic structures. Using an explainable AI (XAI) approach, we show that we can identify the importance of specific spatial regions of a nanophotonic structure for the presence or lack of an absorption peak. Our results highlight that ML strategies can be used for physics discovery, as well as design optimization, in optics and photonics.

## INTRODUCTION

Nanophotonic structures and devices have enabled a broad range of transformative technologies including photonic integrated circuits for optical communication[1-3], and metasurfaces that compactly control the propagation of electromagnetic waves[4-6]. The conventional approach to designing nanophotonic structures is via numerical simulations based on fundamental physical laws (*e.g.*, Maxwell's Equations). This design technique, which we here refer to as 'forward design'[7], is well established, but depends on computationally expensive trial-and-error processes to obtain target functionalities. To address the limitations of forward design, 'inverse design' methods have been developed to generate nanophotonic structures that meet predefined targets[8]. Methodologies such as topology[9-11,31] and adjoint-based optimization[12,13,32] have shown promising results in designing complex structures that deliberately interact with electromagnetic fields, often at sub-wavelength scales, to enable a desired response. While inverse design algorithms can yield high-performance designs that go beyond human intuition, the algorithms can miss globally optimal designs[7,8], produce unstable results[14], are often constrained by long runtimes. Additionally, inverse design methods typically operate as 'black boxes' and cannot explain the underlying relationship between a designed physical structure and its electromagnetic response.

In recent years, machine learning (ML) techniques have emerged as alternate strategies for both forward and inverse design of photonic structures. Tandem neural networks have been used to design multilayer thin films based on target transmission spectra[15], and for spatially complex geometries, generative adversarial networks (GANs) have produced images of structure designs, given an input of desired spectral properties[16-18]. Convolutional neural networks (CNNs) have also been used to map physical geometries to spectral and spatial properties in two- and three-dimensional settings, respectively[33,34]. However, the internal decision models built by these machine learning algorithms are not well-understood; their contents, similar to conventional inverse design approaches, are widely regarded as 'black boxes'[19,20]. This challenge emerges from the fact that supervised ML algorithms, including neural networks, learn by optimizing up to millions of internal variables (weights and biases) to fit the training data[21]. Consequently, it is exceptionally challenging to explain *why* a machine learning algorithm makes one prediction over another. Thus, the lack of explainability is a key limitation for both conventional and ML-based inverse design strategies[35].

In response to the long-standing 'black box' problem of ML, explainable artificial intelligence (XAI) and ML approaches have become a topic of active inquiry. This rapidly developing field aims to analyze and understand

ML models in general, with domain-specific demonstrations of scientific insights that might in turn emerge[36,37]. XAI approaches include sensitivity analysis, Taylor decomposition, deconvolution, guided backpropagation, and layer-wise relevance propagation[38-41]. For image-based classification and regression these methods typically create a salience map (or heatmap) to highlight small portions of the computation that are most relevant to the context at hand, thus explaining the features that contribute most to a model's predictions. While explainability approaches have recently shown promise in better understanding machine learning outcomes in chemical, biological, and physical models[42-45], their use in photonics remains largely unexplored.

Motivated by these developments, in this article, we uncover what a class of machine learning algorithms (CNNs) has learned regarding the underlying physical principles which govern specific classes of nanophotonic structures. CNNs are deep neural networks widely used for image analysis and classification[22]. As shown in Figure 1a, we first created two-dimensional images representing the geometries of metal-dielectric-metal metamaterial resonators. This class of metamaterials was selected due to their ease of fabrication, compact structure, and ability to achieve high absorbance across a broad wavelength range, making them amenable to a wide range of spectral applications while enabling the rapid generation of training data for deep learning. Next, we demonstrate that the CNN can accurately perform forward design by learning the relationships between the metamaterial-structure images and their absorption spectra over mid-infrared wavelengths (Figure 1b). We then use the Deep SHapley Additive exPlanations (SHAP) framework[24] to 'open the black box' and explain the CNN's predictions (Figure 1c). Deep SHAP combines DeepLIFT[46], a previously employed method for decomposing output predictions via backpropagation, with Shapley values, a metric that determines feature relevance, to generate pixel-by-pixel explanation heatmaps[24]. The explanations obtained with Deep SHAP show that the CNN has learned important physical behaviors of the class of metamaterials studied, including the relationships between structural elements and optical responses for both simple and freeform resonator geometries. Our approach uncovers specific geometric contributions to ML predictions of nanophotonic device properties, and thus allow us to both better understand the behavior of complex nanophotonic devices and identify pathways to improved designs (Figure 1d).

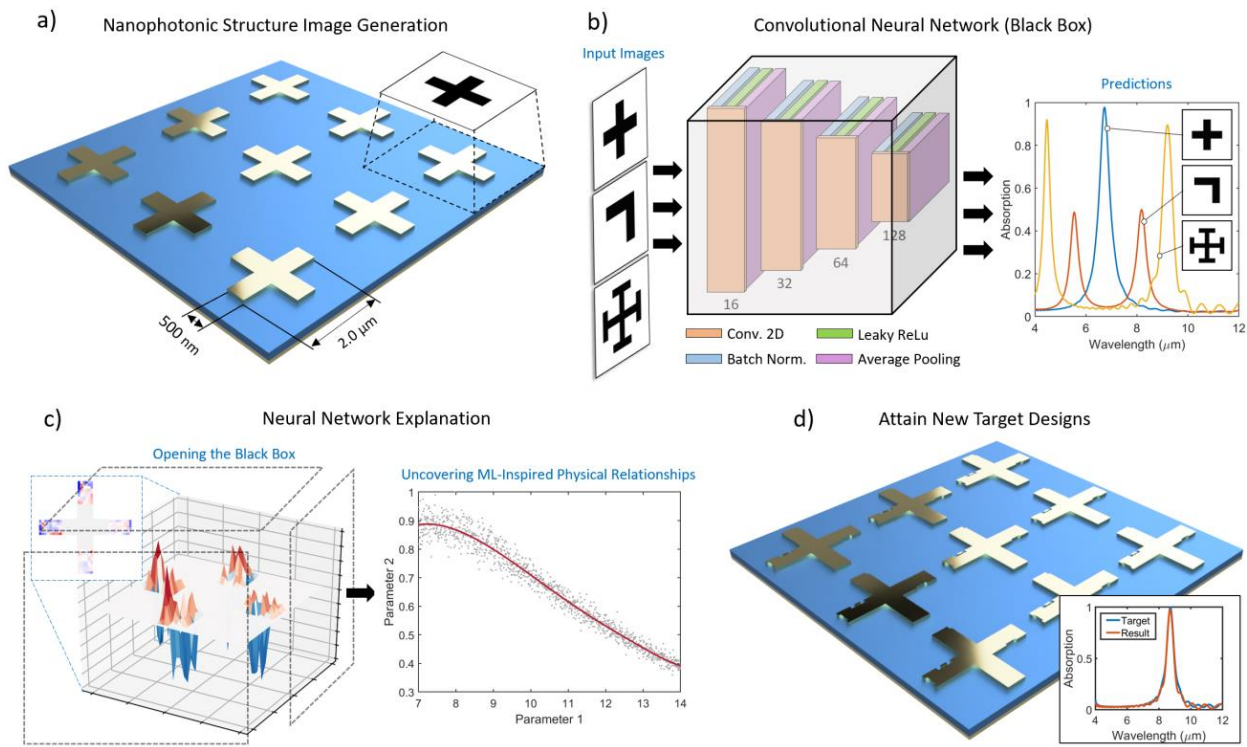

**Figure 1.** (a) Converting 3D metal-dielectric-metal metamaterials into 2D representations for image-based machine learning. (b) Training a convolutional neural network (CNN) to predict the electromagnetic response of input images. (c) Elucidating the underlying physics learned by a CNN by explaining the relationships between structural features and predicted parameters. (d) Leveraging the explained relationships to construct new designs with new target properties.

## RESULTS AND DISCUSSION

**Forward Design Convolutional Neural Network (CNN) Development and Evaluation**

We first developed and trained a CNN for the forward design of nanophotonic structures, such that when it is given an image of a nanophotonic structure as input, it outputs the associated absorption spectrum over a particular wavelength range. To constrain the problem further, we focused on a specific class of nanophotonic structure: metal-insulator-metal (MIM) metamaterials designed to operate at mid-infrared wavelengths[23]. We performed three-dimensional finite-difference time domain (FDTD) simulations of 10,000 unique structures in Lumerical, and generated 10,000 two-dimensional images of the resonator layers and their corresponding absorption spectra. The simulated structures, previously demonstrated in literature to possess selective thermal emissivity over a large bandwidth[23], consist of a 100 nm gold bottom layer, a 200 nm $Al_2O_3$ dielectric middle layer, and a 100 nm gold resonator top layer (within a 3.2 µm × 3.2µm unit cell). The dimensions of the top layer were progressively adjusted for design variation. The designs consist of cross-shapes, box-shapes (hollow and solid), cross-shapes with perpendicular resonators along the arm tips, and the inverted versions of each shape (as shown in Figure S1 of the Supporting Information). The models were then converted into two-dimensional images, and each image was associated with an 800-point vector of absorption values (ranging from 0 to 1) across fixed wavelengths (4 µm to 12 µm). Periodic boundary conditions were applied along the x- and y-planes. Each image was resized to 40 × 40 pixels and converted to grey-scale for ease of training.

After generating the training data, we trained multiple CNN architectures, with 10% of the training dataset used for validation, to determine the optimum hyperparameters. Table S1 presents each of the trained models along with their validation root-mean-square error (RMSE) and training time. Figure 2a shows the predicted output spectra of the CNN when six new and unknown images were used as inputs, as well as the FDTD simulation results corresponding to each image. The simulations were performed by converting the images into the top layer of the MIM structure. We observe through comparison of the simulations and the CNN predictions that the network exhibits a high degree of accuracy in predicting the absorption spectra of a broad range of resonator geometries not present in the training set. The wavelength and amplitude of the predicted resonance peaks are aligned with the simulated peaks (with over 95% mean absolute accuracy). Away from the peaks, we note some minor variation relative to the FDTD simulated results. On average, each prediction was generated in 0.270 ± 0.043 seconds (n = 10), while each simulation took approximately 30 minutes (yielding a 6,500x improvement). The results here demonstrate that the CNN successfully performed the devised forward-design task with high accuracy.

The high accuracy of the CNN's predictions raises an intriguing question: has the CNN, to some extent, learned the physical relationships between the class of nanophotonic structures we explored and their absorption spectra? Normally, the information required to answer this question is embedded within the neural network's many thousands of internal weights and parameters, which is represented by a hierarchy of filters (or neurons). CNNs extract information from images by applying these filters to an input image[28]. The filters are optimized such that the error is minimized when comparing the CNN's output to the target output. Figure 2b shows several examples of the two-dimensional filters that the CNN is composed of (among over 100,000 available filters). The dark squares indicate small weight values and the light squares represent large weight values. As shown in Figure 2c, we can apply these filters to an input image, and capture the CNN's behavior in a series of feature maps. These feature maps can provide insights into the CNN's response to specific areas of the image at any point in the model. For example, in Layer 3, the CNN places more weight on the edges of the cross, while placing less weight on the inside, indicating that generic features such as lines and edges are captured in the initial layers. However, as the input progresses deeper into the model, the feature maps provide progressively less interpretable information. The connection between mapped features and the network's final output remain hard to discern, especially at the deeper layers. Thus, although we can identify the shapes and features extracted by the network's initial layers, it is challenging to synthesize this information into an understandable explanation of the model's decision. Attempting to 'open the black box' in this sense provides limited utility with regards to model interpretation and verification, since analyzing individual filters within a network does not guarantee a coherent explanation for an entire model or even a specific prediction. Furthermore, this form of internal analysis is model specific. Different architectures may yield various feature-map responses in the corresponding layers, leading to inconsistencies in explanations and interpretations.

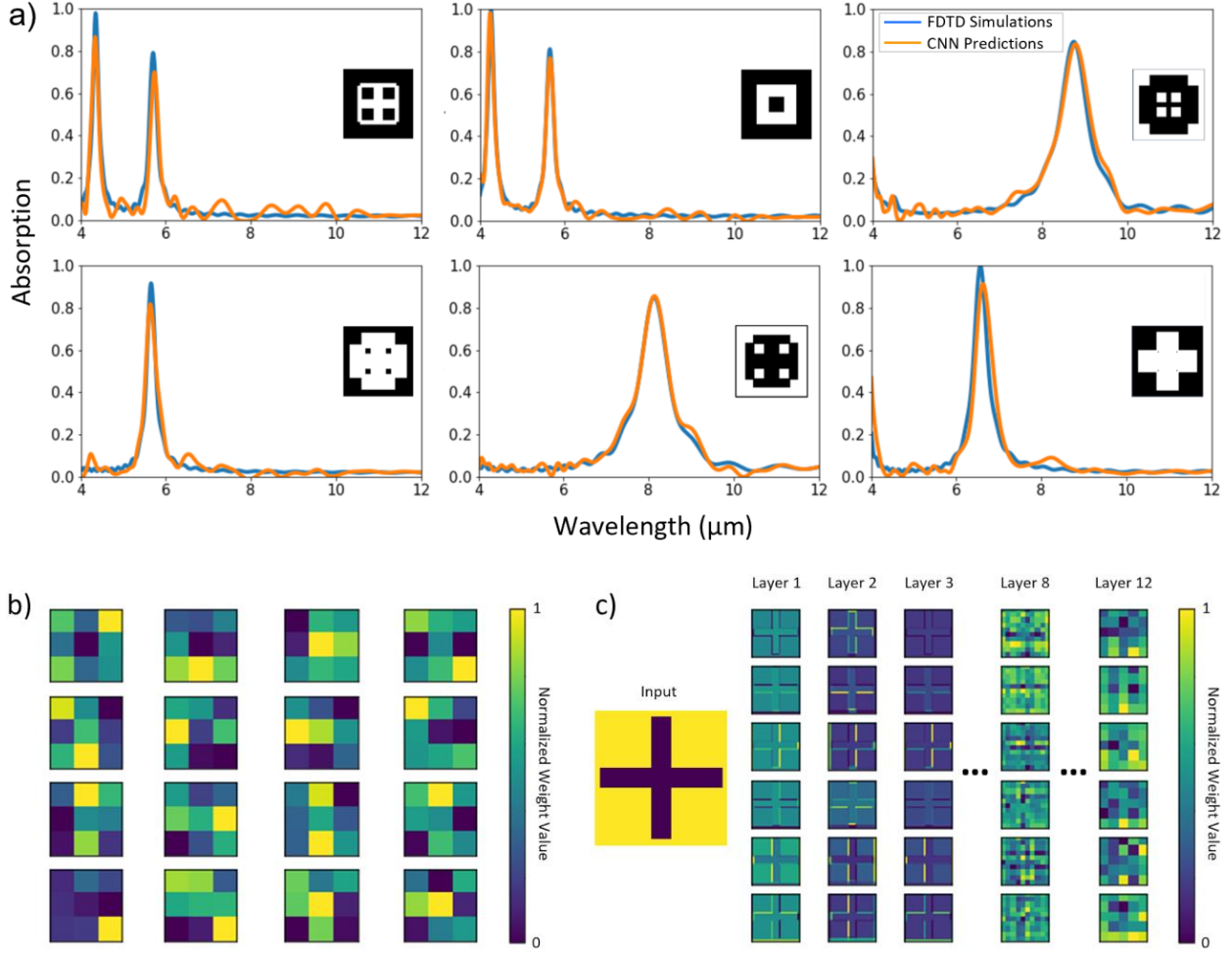

**Figure 2.** (a) CNN-predicted absorption spectra *vs.* FDTD simulations of six new nanophotonic structures (shown in the inset images), revealing the high accuracy of the CNN in performing forward design-based multiphysics structural analysis. (b) Examples of the filters and weights within the CNN. Dark squares represent small weights and light squares represent large weights. (c) Features maps showing what happens inside the model in response to an input image. Generic features such as lines and edges are captured in the initial layers, but the maps are less interpretable as we progress deeper into the model.

**Explaining the CNN's Predictions**

To explain the CNN's behavior and draw useful conclusions from the network's internal model, we instead use the recently developed Deep SHAP method (hereon referred to as SHAP), which attempts to explain model decisions by calculating feature contributions. The methods unified by SHAP are model agnostic and grounded in game theory, leading to more consistent and robust explanations. Instead of compelling the user to analyze thousands of feature maps, SHAP produces a single integrated relevancy-based heatmap that explains a prediction, with results that are output specific and aligned with human intuition, while addressing the previously reported saturation problem and thresholding artifact[46]. With the SHAP values, we can thus explain the contribution of a given geometric feature (represented by its pixels) of a nanophotonic structure to the structure's electromagnetic response at each wavelength. SHAP values are calculated through the following equation:

$$\Phi_i(f,x) = \sum_{z' \subseteq x'} \frac{|z'|!(M-|z'|-1)!}{M!} [f_x(z') - f_x(z' \setminus i)], \tag{1}$$

where, $\Phi_i$ is the SHAP value, $x'$ are simplified inputs that mapped binary values into the original input space ($x$), $M$ is the number of simplified input features, $z'$ is a subset of non-zero indices in $x'$, $f_x(z')$ is a model trained with the feature present, and $f_x(z'\backslash i)$ is a model trained with the feature withheld[24]. The SHAP algorithm captures the effect of withholding a feature, then iterates the computation across all possible subsets $(z' \subseteq x')$. In general, by removing specific features and calculating the change in the output, the algorithm can determine the contribution of these features (positive or negative) towards a specific prediction (*i.e.*, if the change of the output is large, the feature has a large contribution, and vice versa).

We performed SHAP explanations on the CNN model trained on 10,000 images of nanophotonic structures previously described. The explanation is represented as a heatmap, where red pixels represent positive contributions of a base image towards the model's prediction, and blue pixels represent negative contributions. As shown in Figure 3a, SHAP explanation heatmaps were captured at 6.0, 6.4, 6.8, 7.2, 7.6, 8.0, 8.4, and 8.8 μm with single-reference backgrounds (described in the Supporting Information), while the base image possessed a Lorentzian absorption peak at 5.2 μm and arm lengths of 1.4 μm. These explanations reveal the features, or lack thereof, that the CNN deems critical towards achieving an absorption resonance at the designated wavelengths. Specifically, as the resonance wavelength increases, the explanations show regions of blue pixels which gradually migrate from the center of the image to the edges, indicating that starting from the base image, the antenna arm lengths must become longer in order to achieve resonance at larger wavelengths. Conversely, Figure 3b shows that for a base image with longer initial arm lengths (2.9 μm), the arms must become shorter in order to achieve resonance at smaller wavelengths. This behavior is evident from the regions of blue pixels converging towards the center of the image as the resonance wavelength decreases. Both cross-arm tests indicate that the CNN has effectively inferred the relationship between antenna arm length and resonance wavelength.

At the same time, we observe varying degrees of red and blue pixels throughout the explanation heatmaps. For example, on the 8.8 μm explanation with the 5.2 μm base image, there are higher-intensity blue pixels on the top and left arms of the cross, indicating that the CNN weighs each arm differently in determining the resonance wavelength, when in reality, all of the arms are equally important to achieving resonance at the designated wavelength. In addition, the magnitude of the blue pixels are greater towards the edges of the structure, while the remaining areas have red pixels scattered throughout. Both results can be attributed to the filters developed by the CNN during training. CNNs tend to develop edge detection filters, since non-edge patterns (*e.g.*, a patch of black pixels) do not typically provide sufficient information to differentiate discrete objects[22]. Therefore, our CNN was tasked with creating the minimum set of filters that captures the most important features and distinctions (*i.e.*, the cross-arm edges) required to correlate the images to their respective absorption spectra. Naturally, this determines the range of the CNN's feature recognition capabilities and the extent of which it can generalize (or accurately predict new and unknown images), which may be limited to an unknown degree. However, we can alleviate this uncertainty by using the SHAP explanations to identify the sections of the structure that strongly contributed to resonance as well as the sections that contributed only weakly. With this information, we can infer what kind of relationships the CNN learned (or failed to learn), thereby allowing us to determine potential failure modes of the trained model. Thus, in addition to uncovering the physical relationships learned by the CNN, the presented CNN-explanation approach is also effective at determining the limitations and risks associated with a ML model trained on nanophotonic simulations by providing insight into the model's behavior.

To validate the SHAP explanations in their identification of features that contribute to resonance at specific wavelengths, we used the SHAP value heatmaps from Figure 3 to modify the base image, such that the resonance occurs at alternate wavelengths (Figure S2). We compared the SHAP explanations, and the design validations derived from them, against a standard antenna-based analytical relationship between the MIM resonator arm lengths and resonance wavelengths:

$$\lambda = (2n_{eff})L + C. \qquad (2)$$

Here, $\lambda$ is the resonance wavelength, $n_{eff}$ is the effective index of the transverse electric (TE) mode, $L$ is the length of the resonator, and C is a correction phase term[25-27]. The comparison between the SHAP-generated and the FDTD simulated structures demonstrates that the information extracted by the CNN aligns strongly with the physical relationship established in Eqn. 2.

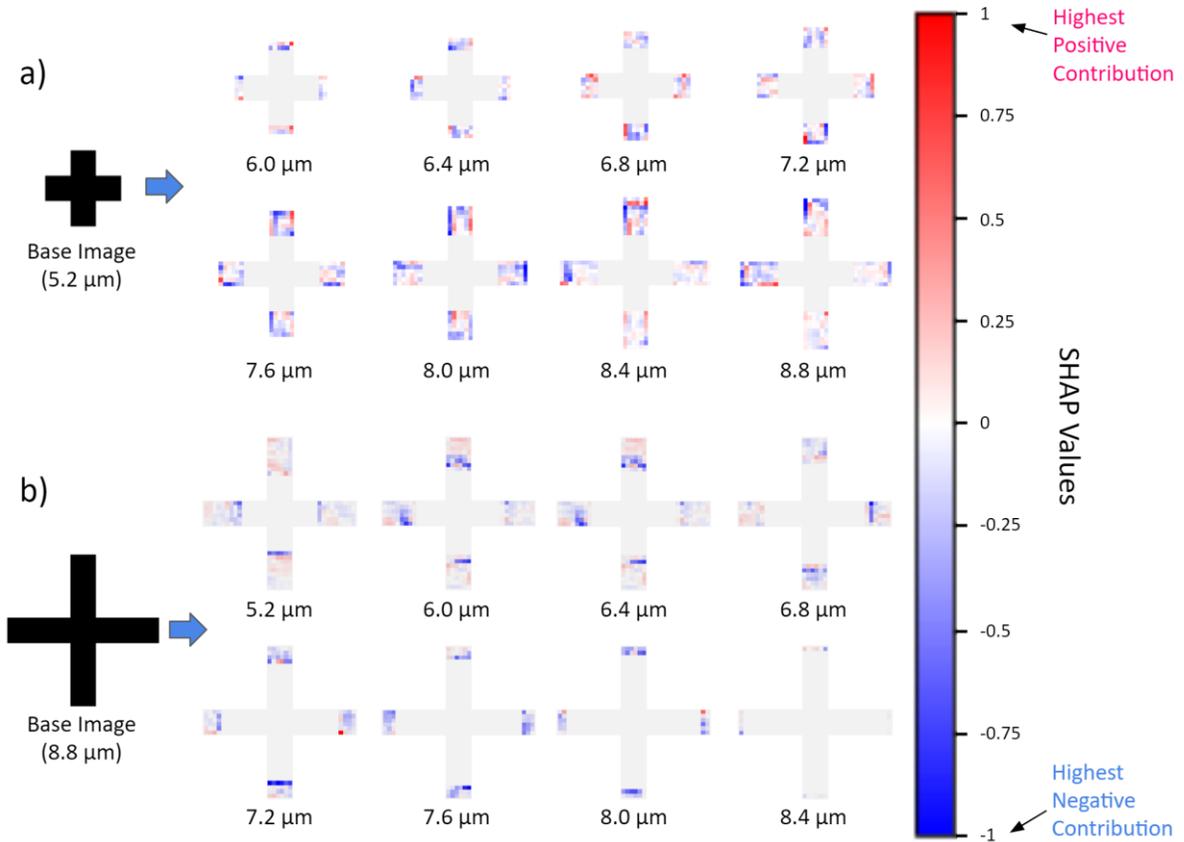

**Figure 3.** SHAP explanations for a (a) 'short-arm' cross (1.4 μm lengths) at increasing resonance wavelengths and a (b) 'long-arm' cross (2.9 μm lengths) at decreasing resonance wavelengths, revealing the CNN learned that the cross-arms must increase to achieve resonance at longer wavelengths and vice versa.

As an additional demonstration of the presented explanation method on complex shape-property relationships, we sought to explain the structural elements that distinguish a structure's response between single and double resonance behaviors within a given bandwidth. Figure 4 presents a series of test cases, where explanations of a dual absorption peak structure (L-shaped) and a single-peak structure (I-shaped) were captured at the peak wavelengths of each structure (marked in Figure 4a). The SHAP explanation heatmaps at the designated wavelengths are shown in Figure 4b, where the I-shaped image was used as the background for the L-shaped image and vice versa. The complete distribution of SHAP values from each heatmap are plotted and quantified in Figure 4c and 4d for the I-shaped resonator and L-shaped resonator, respectively. The inset bar graphs present the average SHAP values across each explanation. From these plots, we observe that at the peak/target wavelengths of the background image, the explanation of the base image at those wavelengths (indicated by the red-dashed boxes in Figure 4b) yield higher-magnitude and more negative SHAP values (blue pixels) than the explanations at non-peak wavelengths. Thus, the results here reveal that the CNN uses the inclusion of the horizontal-bar on the I-shaped structure to determine the presence of two absorption peaks at 5.5 μm and 8.2 μm, while the removal of the bar on the L-shaped structure renders a single peak at 6.3 μm. Using the same design validation strategy from the previous section, we confirmed that the SHAP explanations correctly identified the structural areas that contribute to single and dual resonance (Figure S4).

Furthermore, the explanations provide more granular details on which areas of each nanophotonic structure contributes to each resonance peak. For example, for the dual-peak L-shaped structure, the explanation at each peak (5.5 μm and 8.2 μm) illustrates different red-pixel dominant regions (features contributing to resonance at these wavelengths). We note that the heatmap's spatial distribution bears resemblance to the spatial nature of the resonances on either peak. In particular, the electric field concentrations in these structures vary at different resonance wavelengths (Figure S3). Similar to the electric field of the L-shaped structure at 8.2 μm, SHAP informs us that roughly the entire horizontal-arm length evenly contributes to resonance, while at 5.5 μm, the center of the arm contributes more to the resonance, which aligns with the nature of the E-field distribution for this resonance (Figure S3).

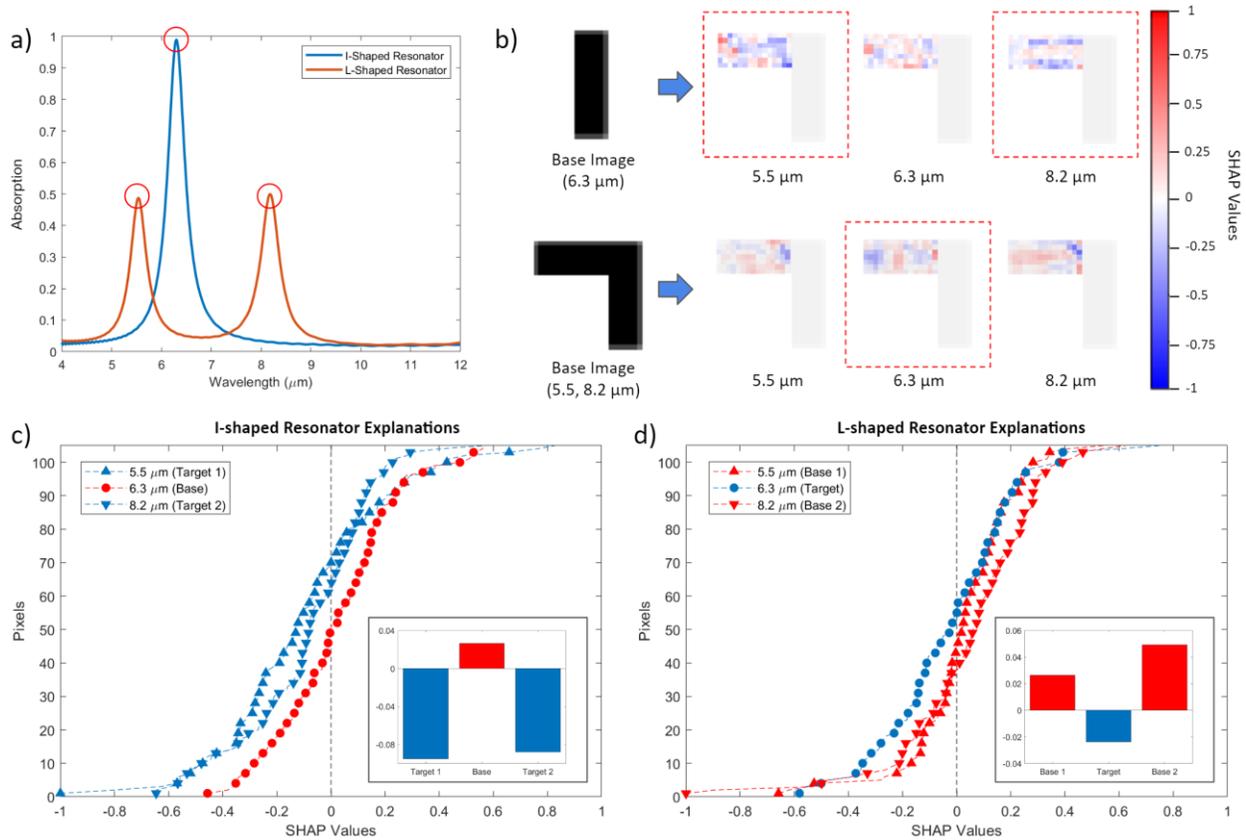

**Figure 4.** (a) Absorption spectra of a single-peak I-shaped resonator and a dual-peak L-shaped resonator. Red circles indicate the resonance wavelengths. (b) SHAP explanations of the resonators at the previously identified resonance wavelengths. Red dashed boxes indicate the explanations for obtaining new target resonance wavelengths of the opposing shape. Distribution of SHAP values across the explanation pixel-maps for the (c) I-shaped resonator and the (d) L-shaped resonator. Inset bar graphs represent the average SHAP values of each explanation, where the negative SHAP values (blue pixels) are dominant on all target explanations.

**Explaining the Response of Complex Freeform Structures**

We next investigate, using SHAP's explanations, the physical insight into the optical response of complex freeform resonator geometries that do not lend themselves to intuitive or previously understood physical models. Figure 5 shows a series of complex freeform metal-insulator-metal metamaterials we examine, as well as the multiple absorption peaks each structure supports. We use SHAP to generate explanations for wavelengths associated with each structure's peaks. In Figure 5, the blue pixels indicate the negative contributions of empty space towards the absorption peak. Therefore, the blue pixels indicate regions of the structure that positively contribute towards the absorption value at a particular wavelengths, while the red pixels indicate the regions that negatively contribute. As seen in Figure 5, the SHAP heatmaps highlight the spatial regions of each freeform shape most responsible for the noted absorption peaks. For example, in Figure 5a, the bottom and left edges of the structure appear to contribute the most to the absorption peak at 7.8 μm, while the top portion contributes to the peak at 8.9 μm. Similarly, in Figure 5c, we observe that the concave part of the structure on the upper-left strongly contributes to resonance at 7 μm, while the same region is not responsible for, and in fact suppresses, the resonance at 4.7 μm. In the case of Figure 5c then, this suggests that both spatial regions are responsible for the combined two peaks observed. Collectively, the negatively and positively contributing regions explain the overall absorption response that is observed at different wavelengths, and establishes a physical picture of the behavior of each freeform shape.

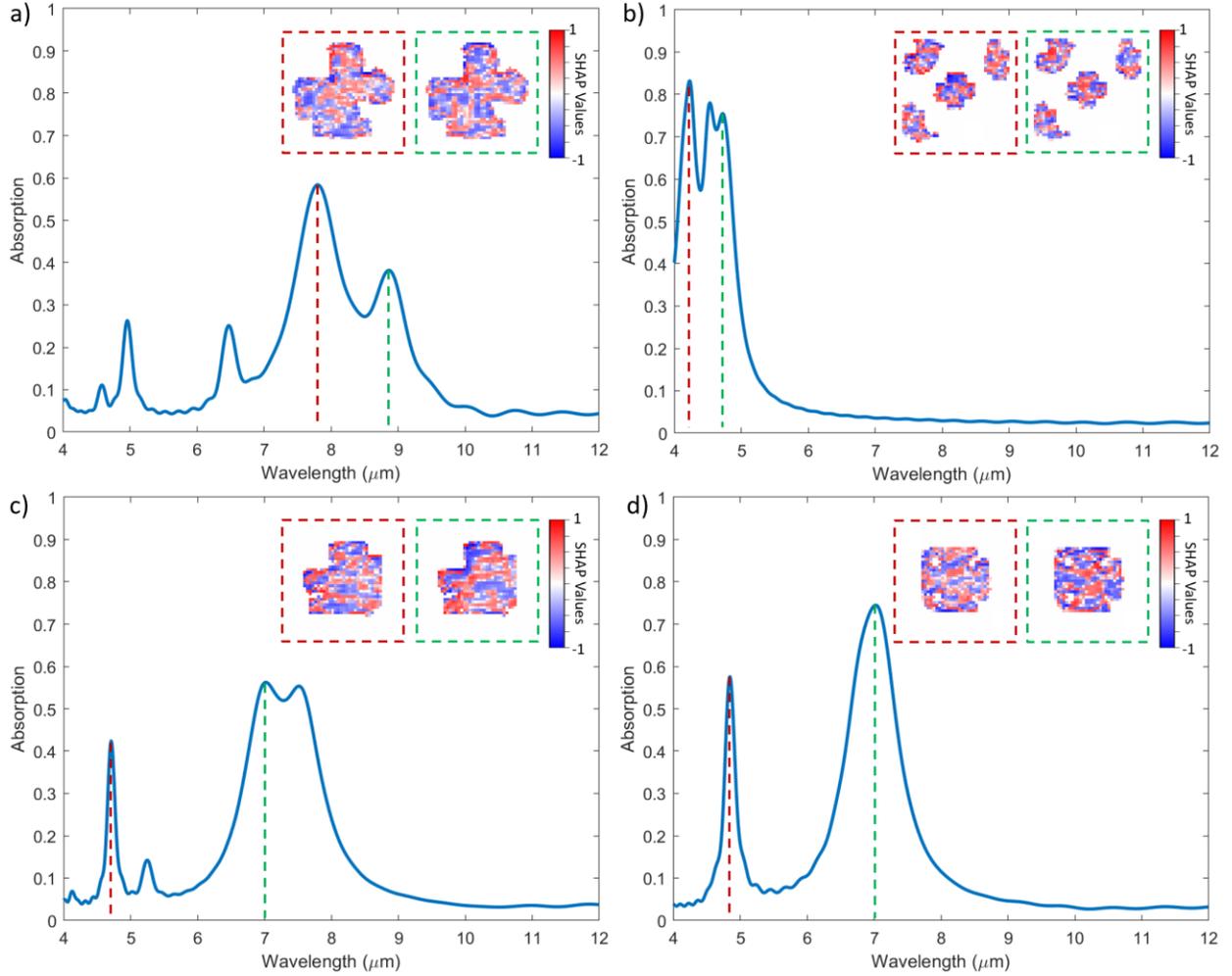

**Figure 5.** (a-d) SHAP explanations for freeform structures at various points of interest, revealing the range of structural elements which contribute to various absorption peaks.

**Validating the Freeform Structure Explanations**

To validate SHAP's explanations of which regions of freeform shapes contribute positively or negatively to the absorption value at a particular wavelength, we modify the original freeform structures based on these contributions, and simulate the modified structures in Lumerical. If SHAP's explanations are accurate, we expect the modified structures to enhance or suppress absorption at a specified wavelength based on the SHAP value for a given spatial region. Focusing the analysis on one resonance wavelength per design, Figure 6a illustrates the explanation of the freeform structure for the absorption peak at 7.8 μm, while 6b, 6c, and 6d show the explanations for their corresponding structures at 4.2 μm, at 7 μm, and 7.1 μm, respectively. After validating the SHAP values, we found that the absorption amplitude of the original structure can be tuned by selectively adding or removing the structural elements informed by SHAP. For example, in Figure 6a, the absorption spectra peaks at 0.6. After generating a structure using primarily blue pixels (shown below the SHAP heatmap and to the right with a blue border), the absorption rose to 0.9. Conversely, using primarily red pixels (which negatively contribute to the absorption value at that wavelength) the generated structure (shown below the SHAP heatmap and to the left with a red border) yields an absorption value near 0.1. We emphasize that this validation step is not necessarily a design strategy for photonic structures. Instead it serves to confirm that the SHAP values at different spatial regions of the freeform structure do in fact correspond to its ability to enhance or suppress absorption at a particular wavelength.

    We note that in deciding which pixels to transform, we used only the largest 95% of the absolute SHAP values to account for noise. We also observe that this absorption enhancement and suppression strategy is consistent

across multiple designs (Figure 6b-d), although the degree of absorption intensity change varies by design, and the resonance wavelengths of the new structures deviate slightly from the original structure. However, from the SHAP heatmaps, we are able to describe the spatial components of an arbitrary structure to discover the regions responsible for absorption at a specific wavelength, and use this information to tune the properties of the metasurface. Thus, the presented XAI method provides an explanation of the behavior of complex structures, where the relationships between structure and property are not readily apparent, and opens the door to new strategies for nanophotonics design.

While promising as a first demonstration, we note some limitations in our current results which are linked to an important current limitation in the Deep SHAP method: its inability to account for feature dependence[29,30]. This in turn could have inhibited the identification of key structural features required for a resonance. Despite the minor discrepancies between the target and resulting resonance wavelengths in our validation studies, the general patterns identified by the explanations still offer significant insights into the features which contribute to resonance; a critical element which was not accessible in previous ML studies pertaining to photonic structures.

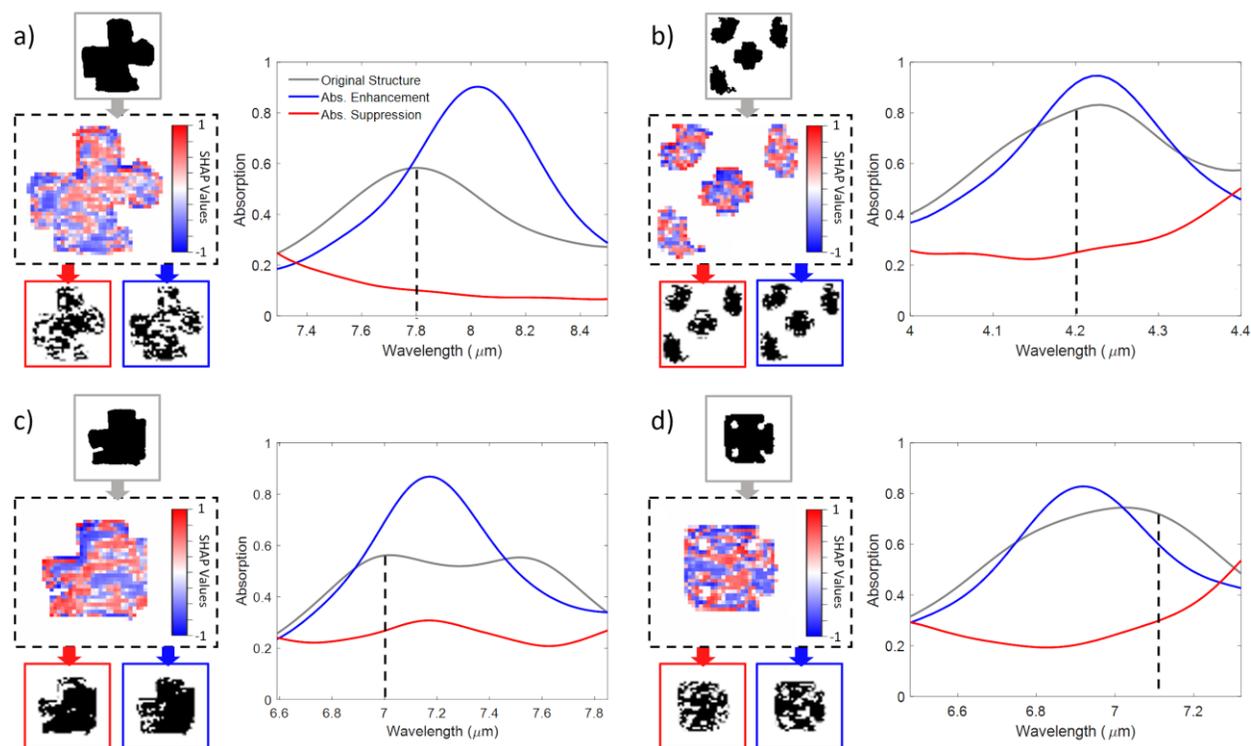

**Figure 6.** Validating the feature contributions highlighted by SHAP. Modifications were made to the freeform structures based on their SHAP-determined feature contributions at (a) 7.8 µm, (b) 4.2 µm, (c) 7 µm, and (d) 7.1 µm. For each structure, the SHAP explanations at the corresponding absorption peaks are shown. The SHAP-determined regions of positive (red pixels) and negative contributions (blue pixels) were used to generate structures which were then simulated using a full-field electromagnetic solver and resulted in absorption peak suppression or enhancement, respectively.

**Conclusion**

In summary, we show that convolutional neural networks can predict the optical properties of nanophotonic structures with remarkable precision, serving as an ultra-fast electromagnetic simulator for constrained domains of structures that also contain valuable information about the behavior of nanophotonic structures. Accordingly, we demonstrated an explanation algorithm (Deep Shapley Additive Explanations, or SHAP) that identifies the contributions of individual image features (on a pixel-by-pixel level) towards each of the network's predictions. The trained CNN predicted the spectra of new and unknown structures with over 95% accuracy, and orders of magnitude faster (~0.3 seconds) than conventional simulation (~30 minutes, yielding a 6,500x improvement). By examining the SHAP explanations, both qualitative and quantitative relationships between structure and spectra can be obtained (*i.e.*, resonator arm length vs resonance wavelength), and the explanations themselves can be used to enable unique design strategies through ML-inspired physics discovery. The explanations also revealed what the CNN did *not* learn, thus exposing potential limitations and risks associated with the trained model. Importantly, the presented explainable artificial intelligence method shows that the patterns and principles encoded within the ML model can be extracted to derive valuable insights into the nanophotonic structure behavior, even in complex freeform structures whose behavior is typically not easy to understand. While we chose to study a class of metamaterial resonators and their corresponding absorption spectra, we emphasize that the approach we have developed is applicable to any class of photonic structure or device for which a sufficiently large training dataset can be assembled by simulations, and any relevant optical device property, including focal depth, field of view and polarization sensitivity. Future studies could thus encompass using emerging explainability algorithms along with the explanation of additional device-property relationships. In the long term, combining explainability with machine learning may enable new discoveries in the physics of highly complex nanophotonic structures, and in turn yield new functionalities and capabilities not possible today.

**Supporting Information**

Nanophotonic structure images from the training dataset, network architecture optimization details, additional SHAP explanation analyses, comparison of SHAP explanations to electromagnetic simulations, and SHAP-informed design validations.

**Notes**

The authors declare no competing financial interest.

**Acknowledgements**

This work was supported by the Sloan Research Fellowship from the Alfred P. Sloan Foundation. This work used computational and storage services associated with the Hoffman2 Shared Cluster provided by UCLA Institute for Digital Research and Education's Research Technology Group.

37. Arrieta, A. B., Díaz-Rodríguez, N., Del Ser, J., Bennetot, A., Tabik, S., Barbado, A., García,S., Gil-López, S., Molina, D., Benjamins, R., et al. (2019). Explainable artificial intelligence (XAI): Concepts, taxonomies, opportunities and challenges toward responsible AI. *Information Fusion* (2019).
38. Simonyan, K., Vedaldi, A. & Zisserman, A. Deep inside convolutional networks: Visualising image classification models and saliency maps. in 2*nd International Conference on Learning Representations, ICLR 2014* - Workshop Track Proceedings (2014).
39. Montavon, G., Lapuschkin, S., Binder, A., Samek, W. & Müller, K. R. Explaining nonlinear classification decisions with deep Taylor decomposition. *Pattern Recognit*. (2017) doi:10.1016/j.patcog.2016.11.008.
40. Zeiler, M. D. & Fergus, R. Visualizing and Understanding Convolutional Networks arXiv:1311.2901v3 [cs.CV] 28 Nov 2013. *Comput. Vision–ECCV 2014* (2014) doi:10.1007/978-3-319-10590-1_53.
41. Bach, S. et al. On pixel-wise explanations for non-linear classifier decisions by layer-wise relevance propagation. *PLoS One* (2015) doi:10.1371/journal.pone.0130140.
42. Lundberg, S. M. et al. Explainable machine-learning predictions for the prevention of hypoxaemia during surgery. *Nat. Biomed. Eng*. (2018) doi:10.1038/s41551-018-0304-0.
43. Ghosal, S. et al. An explainable deep machine vision framework for plant stress phenotyping. *Proc. Natl. Acad. Sci. U. S. A*. (2018) doi:10.1073/pnas.1716999115.
44. Groth, O., Fuchs, F. B., Posner, I. & Vedaldi, A. ShapeStacks: Learning Vision-Based Physical Intuition for Generalised Object Stacking. in Lecture Notes in Computer Science (including subseries Lecture Notes in Artificial Intelligence and Lecture Notes in Bioinformatics) (2018). doi:10.1007/978-3-030-01246-5_43.
45. L. von Rueden, S. Mayer, and et al. Informed machine learning – a taxonomy and survey of integrating knowledge into learning systems. *arXiv preprint* arXiv:1903.12394v2, 2020.
46. Shrikumar, A., Greenside, P. & Kundaje, A. Learning important features through propagating activation differences. in *34th International Conference on Machine Learning, ICML 2017* (2017).
12

**SUPPORTING INFORMATION**

**CNN Training Data and Network Architecture Optimization**

To train our CNN, we performed three-dimensional finite-difference time domain (FDTD) simulations of 10,000 unique nanophotonic structures in Lumerical, and generated 10,000 two-dimensional images of the resonator layers and their corresponding absorption spectra. Figure S1 shows example 2D images of the designs used in the training dataset. The designs consist of cross-shapes, box-shapes (hollow and solid), cross-shapes with perpendicular resonators along the arm tips, and the inverted versions of each shape. To simulate the absorption spectra, these designs were simulated within 3.2 μm × 3.2 μm periodic arrays. Each structure contained a 100 nm gold bottom layer, a 200 nm $Al_2O_3$ dielectric middle layer, and a 100 nm gold top resonator layer with various dimensions. Mesh sizes of 20 nm × 20 nm × 20 nm were maintained across the entire simulation domain, and a plane-wave source at normal incidence was applied across a wavelength window of 4 μm to 12 μm.

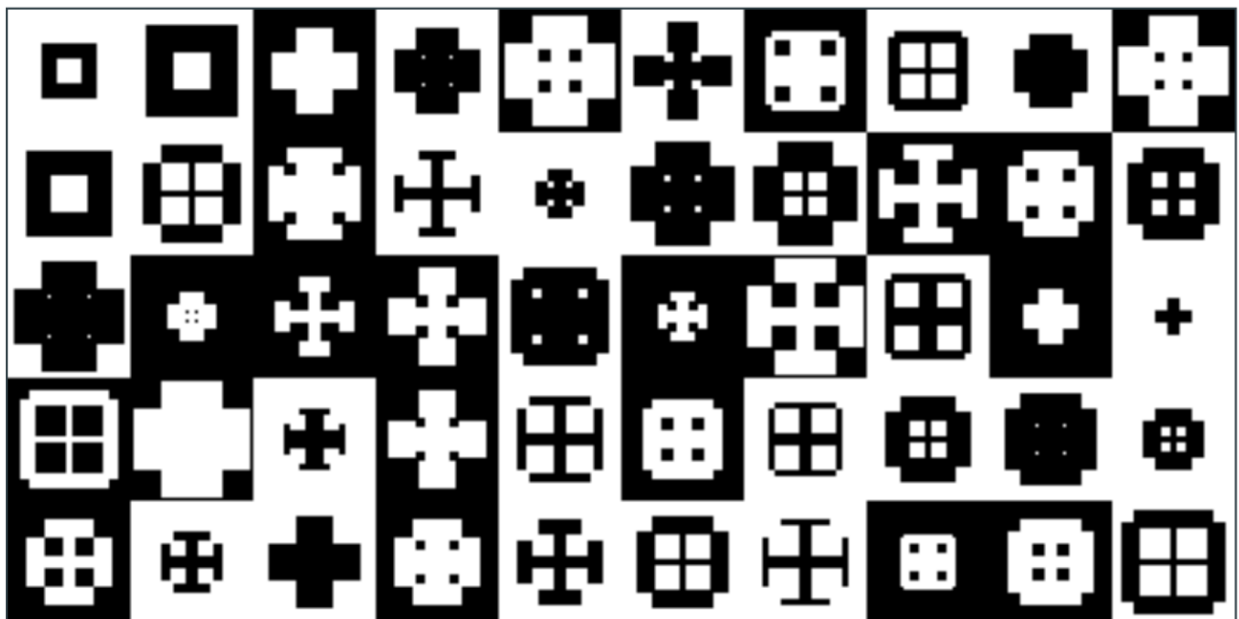

**Figure S1.** A subset of the training data images, consisting of: cross-shapes, box-shapes (hollow and solid), cross-shapes with perpendicular resonators along the arm tips, and the inverted versions of each shape.

The CNN was implemented using TensorFlow and Keras and trained on one Intel Core i5-8600T CPU for 300 epochs. Table S1 presents each of the trained models along with their validation root-mean-square error (RMSE) and training time. Model 1 served as the starting point, which consisted of three convolutional layer-stacks, each proceeding with a batch normalization layer, rectified linear unit (ReLU) activation layer, and average pooling layer (except the final stack). Each convolutional layer used 3 × 3 filters, numbering in 8, 16, and 32 in each subsequent layer. The pooling layer used 2 × 2 windows with a stride of 2. By testing incremental changes to the model (Model 2-8), we determined that a four-stack architecture with leaky ReLU layers trained with the adaptive moment estimation (Adam) algorithm yielded the lowest error (Model 9). In addition, the CNN was trained with a learning rate of 0.001, beta1 of 0.9, beta2 of 0.999, and test dataset of 10%.

|  | **Model 1** |  |  | **Model 2** |  |  | **Model 3** |  |
|---|---|---|---|---|---|---|---|---|
| Layers | Param. | Options | Layers | Param. | Options | Layers | Param. | Options |
| conv2d | 3x3,8 | sgdm | conv2d | 3x3,16 | sgdm | conv2d | 3x3,16 | sgdm |
| ReLU |  | 256 minibatch | ReLU |  | 256 minibatch | leakyReLU |  | 256 minibatch |
| avgPool | 2x2, 2 | 100 epochs | avgPool | 2x2, 2 | 100 epochs | avgPool | 2x2, 2 | 100 epochs |
| conv2d | 3x3,16 |  | conv2d | 3x3,32 |  | conv2d | 3x3,32 |  |
| ReLU |  |  | ReLU |  |  | leakyReLU |  |  |
| avgPool | 2x2, 2 |  | avgPool | 2x2, 2 |  | avgPool | 2x2, 2 |  |
| conv2d | 3x3,32 |  | conv2d | 3x3,64 |  | conv2d | 3x3,64 |  |
| ReLU |  |  | ReLU |  |  | leakyReLU |  |  |
| RMSE | 0.15313 || RMSE | 0.10648 || RMSE | 0.11762 ||
| Time | 63 min || Time | 167 min || Time | 218 min ||
|  | **Model 4** |  |  | **Model 5** |  |  | **Model 6** |  |
| Layers | Param. | Options | Layers | Param. | Options | Layers | Param. | Options |
| conv2d | 3x3,8 | sgdm | conv2d | 3x3,8 | adam | conv2d | 3x3,8 | sgdm |
| ReLU |  | 256 minibatch | ReLU |  | 256 minibatch | ReLU |  | 256 minibatch |
| avgPool | 2x2, 2 | 100 epochs | avgPool | 2x2, 2 | 100 epochs | maxPool | 2x2, 2 | 100 epochs |
| conv2d | 3x3,16 |  | conv2d | 3x3,16 |  | conv2d | 3x3,16 |  |
| ReLU |  |  | ReLU |  |  | ReLU |  |  |
| avgPool | 2x2, 2 |  | avgPool | 2x2, 2 |  | maxPool | 2x2, 2 |  |
| conv2d | 3x3,32 |  | conv2d | 3x3,32 |  | conv2d | 3x3,32 |  |
| ReLU |  |  | ReLU |  |  | ReLU |  |  |
| avgPool | 2x2, 2 |  |  |  |  |  |  |  |
| conv2d | 3x3,64 |  |  |  |  |  |  |  |
| ReLU |  |  |  |  |  |  |  |  |
| avgPool | 2x2, 2 |  |  |  |  |  |  |  |
| conv2d | 3x3,128 |  |  |  |  |  |  |  |
| ReLU |  |  |  |  |  |  |  |  |
| RMSE | 0.13289 || RMSE | 0.11497 || RMSE | 0.16737 ||
| Time | 87 min || Time | 77 min || Time | 58 min ||
|  | **Model 7** |  |  | **Model 8** |  |  | **Model 9** |  |
| Layers | Param. | Options | Layers | Param. | Options | Layers | Param. | Options |
| conv2d | 3x3,8 | sgdm | conv2d | 3x3,8 | sgdm | conv2d | 3x3,16 | adam |
| ReLU |  | 256 minibatch | ReLU |  | 128 minibatch | leakyReLU |  | 128 minibatch |
| avgPool | 2x2, 2 | 300 epochs | avgPool | 2x2, 2 | 100 epochs | avgPool | 2x2, 2 | 300 epochs |
| conv2d | 3x3,16 |  | conv2d | 3x3,16 |  | conv2d | 3x3,32 |  |
| ReLU |  |  | ReLU |  |  | leakyReLU |  |  |
| avgPool | 2x2, 2 |  | avgPool | 2x2, 2 |  | avgPool | 2x2, 2 |  |
| conv2d | 3x3,32 |  | conv2d | 3x3,32 |  | conv2d | 3x3,64 |  |
| ReLU |  |  | ReLU |  |  | leakyReLU |  |  |
|  |  |  |  |  |  | avgPool | 2x2, 2 |  |
|  |  |  |  |  |  | conv2d | 3x3,128 |  |
|  |  |  |  |  |  | leakyReLU |  |  |
| RMSE | 0.097562 || RMSE | 0.14086 || RMSE | 0.07709 ||
| Time | 229 min || Time | 42 min || Time | 340 min ||

**Table S1.** Table of trained CNN architectures and the corresponding RMSE values.

**SHAP Explanation Validation for Single-Resonance Structures**

The DeepExplainer module from the Deep SHAP Python library was used to explain the predictions of the CNN. To generate SHAP values for deep learning models, the SHAP algorithm approximates the conditional expectations of SHAP values using a selection of background samples. The background dataset is used to determine the impact of a feature by replacing the feature with values from the background. In doing so, the algorithm can simulate 'missing' features and calculate the impact on the model output[1]. To minimize the noise that was generated by the SHAP explanations, we performed the SHAP explanations under single-reference background conditions. For the single-reference background, we used an image with a specific absorption peak as the background (*e.g.*, 8.0 µm), and captured the explanation at this peak wavelength. This process was repeated for all target wavelengths.

Following the generation of the explanations, the explanations were used for design validations towards target resonances. Validation was performed by converting the blue pixels in the heatmaps to black pixels on the base image. Figure S2b and S2c show the spectra of the validated structures and the original FDTD simulated structures, respectively. These validated designs were then compared with the corresponding FDTD simulated background images (as shown in Figure S2a) to ensure that the CNN learned the relationship between cross-arm length and resonance wavelength. In Figure S2d, the resonant wavelengths at peak absorption and the antenna arm lengths of both sets of structures are plotted (with linear fits of R$^2$=0.998). The FDTD-simulated structures yield an $n_{eff}$ of 1.13 and C of 2.21, while the SHAP-validated structures display an $n_{eff}$ of 1.15 and C of 2.10, yielding an $n_{eff}$ error of 1.8% and a C error of 4.9%.

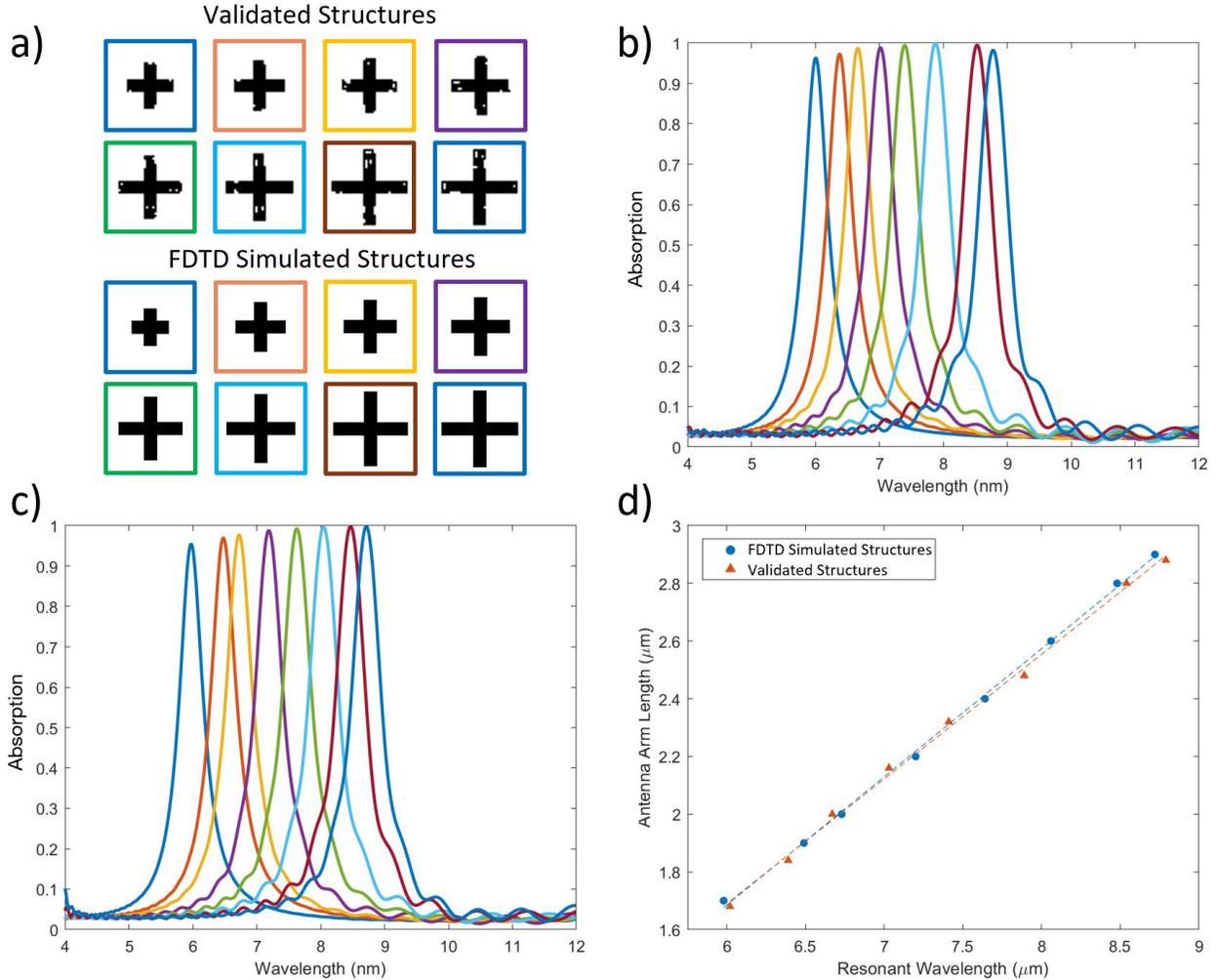

**Figure S2.** (a) Images of the SHAP-validated and FDTD simulated structures. The absorption spectra for the corresponding (b) validated and (c) simulated structures. Image-border colors correspond to the plot colors. (d) Comparison of the physical relationship between antenna arm length and resonant wavelength for the two sets of structures (linear fit of plots shown with R$^2$=0.998).

**SHAP Explanation Validation for Multi-Resonance Structures**

Figure S3 shows the electric field profiles of various MIM structure designs. We note that the SHAP explanation heatmap's spatial distribution bears resemblance to the spatial nature of the resonances on either peak. In particular, the electric field concentrations in these structures vary at different resonant wavelengths.

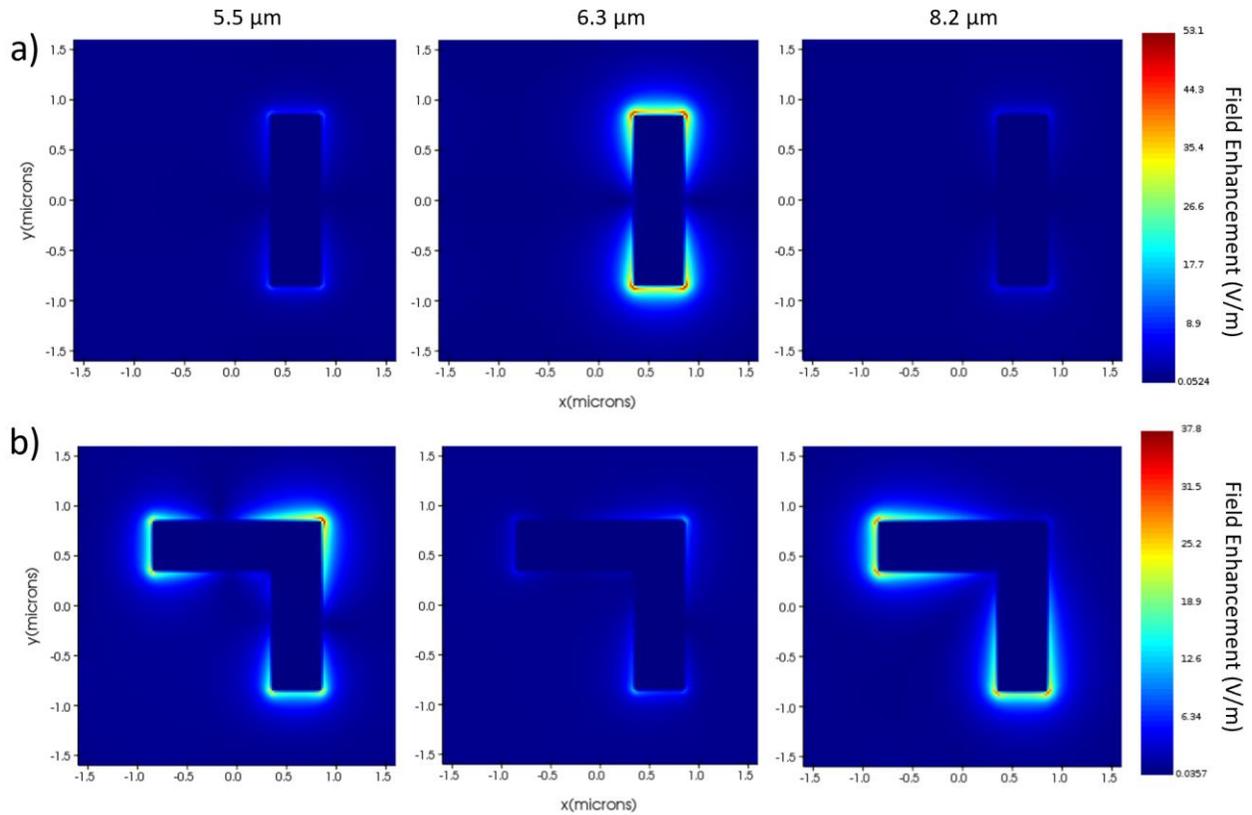

**Figure S3.** The electric field simulation profile of (a) an L-shaped resonator and (b) an I-shaped resonator at resonance wavelengths of 5.5 µm, 6.3 µm, and 8.2 µm.

    We performed the same design validation method from the previous section to assess the accuracy of the SHAP explanations for complex, multi-resonance structures. In Figure S4a, the L-shaped structure was validated by utilizing the explanation generated at 6.3 µm, then converting all of the blue pixels to the opposite state on the original image. The resulting structure exhibited a single absorption peak of approximately 0.9 at 5.4 µm. Using the same approach, we attempted the reverse scenario of generating a dual-peak structure from a single-peak structure (Figure S4b). We leveraged the explanation from one of the dual-peak wavelengths (as either wavelength resulted in negligible differences) and applied it to the design validation process. The validated structure possessed an absorption peak of ~0.6 at 4.8 µm and ~0.48 at 6.9 µm. The design validation studies demonstrate that complex spectral targets can be met by converting the pixels identified by the SHAP heatmaps, and thus that the heatmaps themselves reveal useful information about the relationship between geometric features and their electromagnetic response. In the first case, by focusing the image conversion process on the explanations of a single target wavelength, we converted a dual-peak structure into a single-peak structure. In the second case, the single-peak structure was converted into a dual-peak structure by using the SHAP values of two target wavelengths.

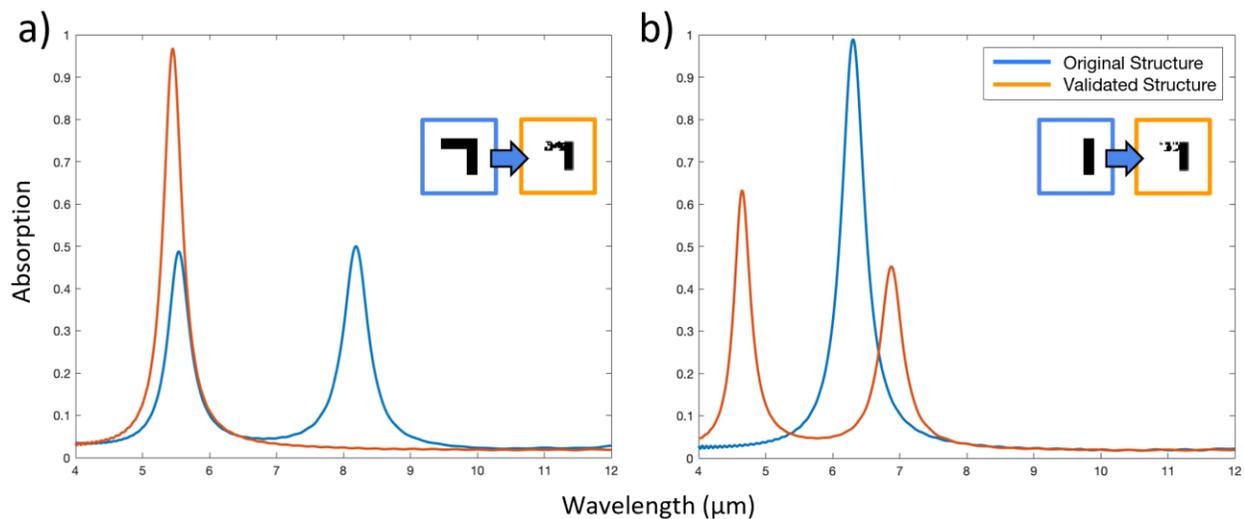

**Figure S4.** SHAP-validation for a (a) single-peak structure and a (b) dual-peak structure by utilizing the SHAP values at targeted resonance wavelengths for image conversion.

**Supporting Information References**